# Jet-gas interactions in z~2.5 radio galaxies: evolution of the ultraviolet line and continuum emission with radio morphology

A. Humphrey,[1,2]♠ M. Villar-Martín,[3] R. Fosbury,[4] J. Vernet,[5] S. di Serego Alighieri[6]

[1]*Department of Physics, Astronomy and Mathematics, University of Hertfordshire, Hatfield, Hertfordshire, AL10 9AB, UK*
[2]*Instituto de Astronomía, Universidad Nacional Autónoma de México, Apartado Postal 70-264, 04510 México, DF, Mexico*
[3]*Instituto de Astrofísica de Andalucía (CSIC), Aptdo. 3004, 18080 Granada, Spain*
[4]*Space Telescope - European Coordinating Facility, Karl Schwarzschild Str. 2, D-85748 Garching bei München, Germany*
[5]*European Southern Observatory, Karl Schwarzschild Str. 2, D-85748 Garching bei München, Germany*
[6]*INAF - Osservatorio Astrofisico di Arcetri, Largo E. Fermi 5, I-50125, Firenze, Italy*

**ABSTRACT**

We present an investigation into the nature of the jet-gas interactions in a sample of 10 radio galaxies at $2.3 < z < 2.9$ using deep spectroscopy of the UV line and continuum emission obtained at Keck II and the Very Large Telescope. Kinematically perturbed gas, which we have shown to be within the radio structure in previous publications, is always blueshifted with respect to the kinematically quiescent gas, is usually spatially extended, and is usually detected on both sides of the nucleus. In the three objects from this sample for which we are able to measure line ratios for both the perturbed and quiescent gases, we suggest that the former has a lower ionization state than the latter.

We propose that the perturbed gas is part of a jet-induced outflow, with dust obscuring the outflowing gas that lies on the far side of the object. The spatial extent of the blueshifted perturbed gas, typically ~35 kpc, implies that the dust is spatially extended at least on similar spatial scales.

We also find interesting interrelationships between UV line, UV continuum and radio continuum properties of this sample.

## 1. INTRODUCTION

Powerful radio galaxies are important cosmological probes for understanding when, and how, the most massive galactic systems formed. They are believed to be hosted by massive ellipticals (e.g. McLure et al. 1999) residing within cluster or group environments (e.g. Pentericci et al. 2000a) and, therefore, represent beacons for finding such systems. Moreover, there is a growing body of evidence for a symbiosis between the evolution of the host galaxies and the nuclear and radio jet activity: the redshift evolution of the co-moving number density of radio sources shows a marked similarity to the global star formation rate of the universe (e.g. Madau et al. 1996; Dunlop & Peacock 1990); and properties of the stellar bulge are apparently correlated with the mass of the central super-massive black hole (e.g. Magorrian et al. 1998;

---

♠ email: ahumphre@astroscu.unam.mx

McLure & Dunlop 2002).

There is now considerable evidence to suggest that the radio jet can undergo strong interactions with the ambient ISM; such jet-gas interactions[*] can affect profoundly many of the observed properties of the hosts and environments of radio galaxies. The extended emission line regions (EELR hereinafter) of some radio galaxies show highly perturbed gas kinematics (FWHM ~1000 km s$^{-1}$); since these perturbed kinematics are usually associated with bends and hotspots in the radio structure (e.g. van Breugel et al. 1985, 1986; Baum, Heckman & van Breugel 1992), it is thought that they result from jet-gas interactions.

The degree to which shocks, driven by jet-gas interactions, contribute to the ionization of the EELR has been the subject of much debate in the literature (e.g Tadhunter 2002). While photoionization must contribute substantially in the majority of cases (e.g. Robinson et al. 1987; Villar-Martín, Tadhunter & Clark 1997), there is mounting evidence in favour of the notion that, in at least some radio galaxies, shocks contribute significantly to the ionization of the EELR. Among the most compelling evidence for this is the discovery by Tadhunter et al. (2000) of emission line filaments in Coma A and 3C 171 that lie outside any plausible illumination cone and, in the former object, circumscribe the radio lobes. Close morphological association between the radio and line emission also provides strong evidence for shock ionization (e.g. Clark et al. 1998; Solorzano-Iñarrea, Tadhunter and Bland-Hawthorn 2002). Further evidence for ionization by shocks comes from the measurement, in perturbed gas associated with the radio hotspots, of electron temperatures and line ratios which cannot be readily explained by photoionization models (e.g. Villar-Martín et al. 1999a). A variety of other results have been cited as providing evidence, though somewhat circumstantial, for shock ionization including: relatively high electron temperatures (e.g. Tadhunter, Robinson & Morganti 1989), and line ratios that are more consistent with shock-ionization than with photoionization (e.g. Solorzano-Iñarrea, Tadhunter & Axon 2001).

There is also evidence to suggest that the morphology of the radio emission is linked to that of UV-optical continuum. The latter, which tends to be closely aligned with the radio axis at z>0.6 (e.g. Chambers, Miley & van Breugel 1987; McCarthy et al. 1987), has been shown by Best, Longair & Röttgering (1996) to undergo an apparent evolution with the former: smaller radio sources tend to exhibit knottier, brighter continuum structures, which are more closely aligned with the radio axis, than shown by larger radio sources.

Our group has been carrying out a programme of quantitative spectroscopy of radio galaxies in the early universe, with the principal goals of understanding the formation and evolution of massive elliptical galaxies, and the way in which the nuclear and radio jet activity affects these processes (see Fosbury et al. 2003). We have targeted sources at z~2.5 in the hope of catching their hosts in the act of formation.

Detailed modelling of the spatially integrated properties of z~2.5 radio galaxies (HzRG hereinafter) has revealed that the UV continuum emission is generally dominated by nuclear light scattered by highly clumped dust (Vernet et al. 2001). The spatially integrated UV emission line spectra have also been studied and demonstrate that high metallicities are common in the EELR of z~2.5 radio galaxies (Vernet et al. 2001; Villar-Martín et al. 2001).

In two recent papers (Villar-Martín et al. 2002, 2003; hereinafter VM2002 and VM2003), we examined the kinematic properties of the extended emission line regions along the radio axis of 10 z~2.5 radio galaxies and found, in most objects, (i) perturbed gas typically with high surface brightness confined by the radio structures (FWHM and velocity shifts >1000 km s$^{-1}$), and (ii) quiescent[†] low surface brightness haloes across the entire observed extent of the EELR (FWHM and velocity shifts of several hundred km s$^{-1}$). We proposed that jet-gas interactions are responsible for the kinematic properties of the perturbed gas, and that the quiescent gas is the AGN-photoionized, non-shocked halo.

Previous investigations by our group have concentrated on understanding the spatially integrated properties of the continuum and line emission, or on searching for kinematically quiescent haloes. In this paper we focus on understanding the possible impact of the radio jets on the ionization, kinematic and continuum properties of z~2.5 radio galaxies.

This paper is organized as follows. The data, and the analysis thereof, is described in §2. General results, and results concerning individual objects, are presented in §3. The nature of the quiescent and perturbed gases is discussed, and a scenario to explain the results is developed, in §4. Finally, the conclusions are summarized in §5.

---

[*] The term 'jet-gas interaction' is used hereinafter as a description of interactions between the radio emitting plasma and the interstellar or intracluster medium.

[†] We use the term 'quiescent' to mean 'in a quasi-stable dynamical configuration in a galaxy', that is, within a factor of 2 of the rotational velocity.

Throughout this paper we assume a flat universe with $H_0 = 65$ km s$^{-1}$ Mpc$^{-1}$, $\Omega_\Lambda = 0.7$ and $\Omega_M = 0.3$. This gives a spatial scale of 8.9 to 7.7 kpc arcsec$^{-1}$ for the range in redshift of 2.3 to 3.6 covered by this sample.

## 2. DATA AND ANALYSIS

The data-set used for this investigation comprises long-slit spectropolarimetry, obtained at the Keck II telescope using the Low Resolution Imaging Spectrometer in polarimetry mode (LRISp; Oke et al. 1995; Goodrich, Cohen & Putney 1995), for a sample of nine radio galaxies selected from the Leiden ultra-steep spectrum compendium ($F_\nu \sim \nu^\alpha$ where $\alpha < -1.0$; e.g. Röttgering et al. 1995) and the 4C ultra-steep spectrum compendium (e.g. Chambers & Miley 1989). These compendia and, therefore, our sample are representative of high radio power and optically bright radio galaxies. We selected sources with $2.3 < z < 2.9$ so that Ly$\alpha$ $\lambda 1216$ could be observed at $\lambda > 4000$Å where the sensitivity of LRIS is high, and so that CIII] $\lambda\lambda 1907, 1909$ could also be observed. For all objects, the slit was aligned with the radio axis as determined by Röttgering et al. (1994) and Carilli et al. (1997). The FWHM of the instrumental profile (IP hereinafter) ranges between 500±60 and 780±70 km s$^{-1}$ at the redshifted wavelength of HeII λ1640, measured from unblended sky lines (see table 1). Vernet et al. (2001), VM2002 and VM2003 provide further details of these data and their reduction.

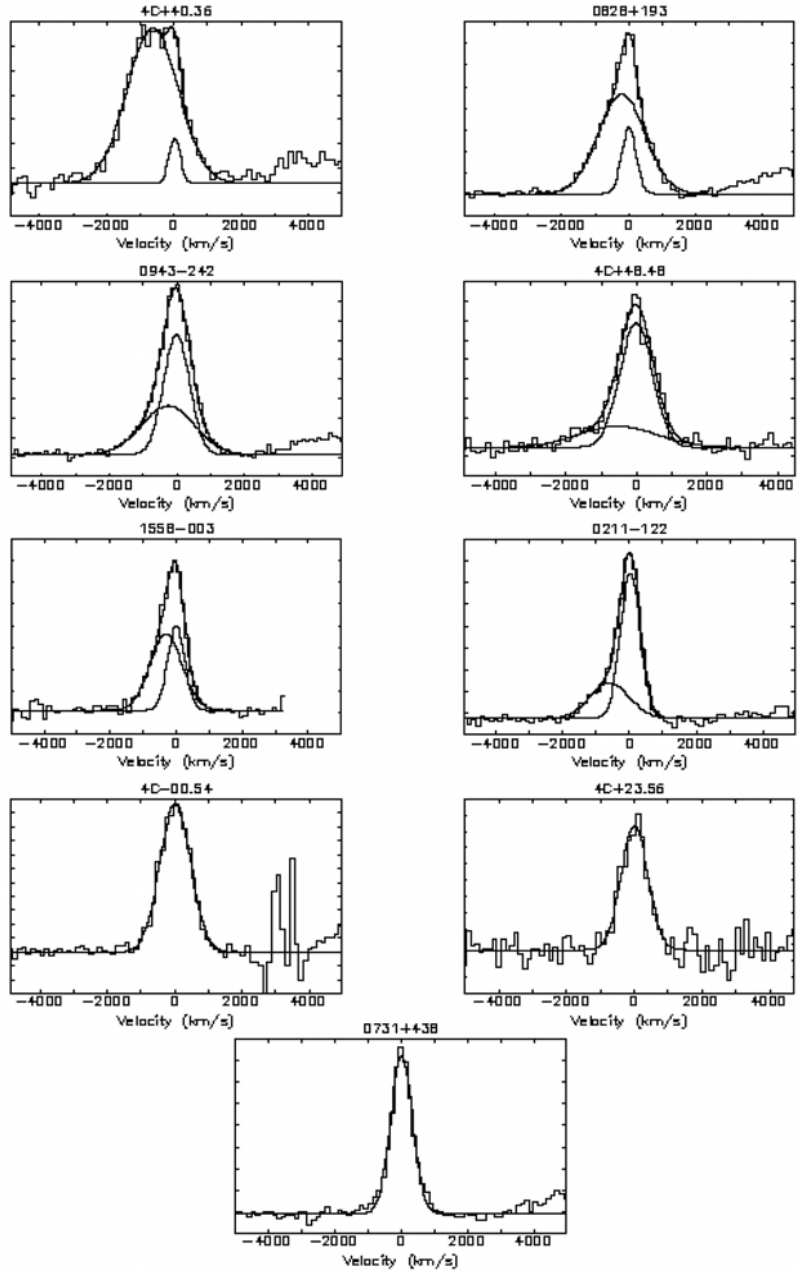

**Figure 1:** The HeII velocity profiles for the HSBR, organized such that objects showing the strongest evidence for jet-gas interactions (see text) are nearer the top. Also shown are the Gaussian fits. The vertical axis – flux density – is in arbitrary units.

**Table 1.** The kinematic properties of the HeII emission from the HSBR, along with the salient radio properties. [1] Name of the source [2] Redshift of the source [3] Instrumental profile (IP) at the wavelength of redshifted HeII, measured from unblended sky lines [4] FWHM of the quiescent component to HeII in the NUC aperture, with the instrumental profile subtracted in quadrature; in the case of 2104-242, the FWHM of the NE emission component measured by Overzier et al. (2001) is used [5] FWHM of the perturbed component to HeII in the NUC aperture, with the instrumental profile subtracted in quadrature [6] Velocity shift of the perturbed component to HeII relative to the quiescent component to HeII [7] Flux of the perturbed component to HeII divided by the flux of the quiescent component to HeII [8] Global FWHM of HeII measured in the spatially integrated spectrum (taken from Vernet et al. 2001; 2104-242: Overzier et al. 2001) [9] Properties of the radio structures: H. Hotspots are coincident with the HSBR B. A bend in the radio jet is coincident with the HSBR N. Radio image shows no indication of jet-gas interactions within the HSBR [10] The projected diameter of the radio structure. NOTE – the quiescent component of 4C+40.36, 0828+193 and 1558-003 is narrower than the IP and, therefore, only upper limits to their FWHM are listed. The correlation between HeII P/Q and the global FWHM of HeII (see text) has a Spearman rank correlation coefficient of >94 per cent, or 100 per cent when the data are binned by a factor of 2 in P/Q or FWHM.

| Source | z | IP (km s$^{-1}$) | Quiescent FWHM (km s$^{-1}$) | Perturbed FWHM (km s$^{-1}$) | $v_s$ (km s$^{-1}$) | HeII P/Q | Global FWHM (km s$^{-1}$) | Radio Properties | Radio Size (kpc) |
| --- | --- | --- | --- | --- | --- | --- | --- | --- | --- |
| [1] | [2] | [3] | [4] | [5] | [6] | [7] | [8] | [9] | [10] |
| 4C+40.36 | 2.265 | 550±70 | <500 | 1710±90 | -650±50 | 16±4 | 1655 | H | 32 |
| 0828+193 | 2.572 | 540±60 | <500 | 1570±80 | -290±30 | 5.1±0.6 | 1366 | B | 98 |
| 1558-003 | 2.527 | 570±60 | <500 | 990±70 | -310±110 | 1.7±0.7 | 865 | N | 71 |
| 0943-242 | 2.922 | 500±60 | 750±130 | 1760±100 | -250±50 | 0.9±0.3 | 1042 | H | 29 |
| 4C+48.48 | 2.343 | 610±70 | 820±160 | 2200±310 | -200±120 | 0.5±0.1 | 922 | H B | 110 |
| 0211-122 | 2.340 | 520±70 | 470±100 | 1290±160 | -620±160 | 0.5±0.1 | 626 | B | 134 |
| 4C-00.54 | 2.360 | 780±70 | 680±170 | - | - | ~0 | 788 | N | 189 |
| 4C+23.56 | 2.479 | 720±60 | 550±130 | - | - | ~0 | 749 | N | 411 |
| 0731+438 | 2.429 | 560±64 | 520±110 | - | - | ~0 | 560 | N | 84 |
| 2104-242 NE | 2.491 | 310±30 | 450 | - | - | ~0 | 450 | N | 177 |

In addition, we use a long-slit spectrum of the z=2.49 radio galaxy 2104-242, obtained at the Very Large Telescope Antu telescope using the Focal Reducer/Low Dispersion Spectrograph (FORS1; Appenzeller et al. 1992). The slit was 1" wide and oriented at a position angle of 18° and thus encompassed the brightest emission components. The FWHM of the IP is 310±30 km s$^{-1}$ at the redshifted wavelength of HeII λ1640. Overzier et al. (2001) provide further details of these data.

In order to obtain a 'standard' measure of the surface brightness of the perturbed and quiescent components across the high surface brightness regions (HSBR), Gaussians were fitted to the HeII λ1640 emission from the central 2.1" (18.8 to 18.0 kpc). This line in particular was used because it is a non-resonant, single recombination line and is relatively insensitive to a wide variety of physical conditions, although it is sensitive to the shape of the ionizing continuum. Note that using CIV in place of HeII does not affect significantly the results of this paper. For all objects, a single Gaussian was initially tried, and only when the observed emission profile provided clear evidence to suggest the need for a second component was one added. The resulting fits are in good agreement with those of VM2002 and VM2003. The assumption that the perturbed component can be represented by a single Gaussian is supported by higher-resolution spectroscopy of HeII and other UV lines in 0943-242 (Binette et al. 2000; see also Jarvis et al. 2003), which shows a smooth, Gaussian-like velocity profile.

It was necessary to take approaches of a somewhat more ad hoc nature to obtain line ratios for the perturbed and quiescent kinematic components. The first of these was to identify regions where both kinematic components are present and fit these two components using a Gaussian for each. The density sensitive CIII] doublet was assumed to be in its low density limit (<1000 cm$^{-3}$), corresponding to [CIII] λ1907/CIII] λ1909 = 1.5. The NV and CIV emission doublets were assumed to have optically thick intensity ratios of 1:1, though the fits are not significantly affected if the optically thin ratios are instead used. For both doublets the sub-components were assumed to have equal FWHM. The flux, velocity and width of each Gaussian, and also the continuum level, were free parameters. Fits were considered to be consistent if the perturbed and quiescent components have consistent velocities and widths, within the uncertainties, for at least three emission lines. Only at a few positions of two objects, namely 0828+193 and 4C+40.36, was it possible to use this particular method. (In other objects the fits, especially to the perturbed component, were of low quality.) For 4C+40.36, Lyα, CIV, HeII and CIII] were fitted consistently. In the case of 0828+193, it was possible to obtain consistent fits to NV, CIV, HeII and CIII]. However, Lyα is strongly affected by absorption in this object (van Ojik et al. 1997) and, therefore, it was not possible to obtain a fit consistent with those obtained for the other lines. There is no evidence for resonance scattering in the observed CIV or NV profiles, and the consistency between lines (i.e. NV, CIV, HeII and CIII]) of the fits supports the notion that resonant scattering does not strongly affect the two aforementioned lines. (Resonant scattering would affect the velocity profile of the resonance lines due to the scattering of energy into the wings and the introduction of absorption features.)

The second approach was to spatially isolate regions that are dominated by either the perturbed or quiescent component, and measure line ratios therefrom. Although line ratios measured in this way are strictly not those of the perturbed or quiescent component, they can be considered to be approaching the true values. These measurements are given in Humphrey (2004) and Humphrey et al. (2006, in preparation), where we present for this sample detailed analysis and modelling of the spatial variation of the line ratios. In the cases of 4C+40.36 and 0828+193, the ratios derived in this manner are consistent with those determined from the fits.

## 3. RESULTS

### 3.1 Notes on individual objects

The properties that are evidential for, or against, the occurrence of jet-gas interactions are described for each object in this sample. We recapitulate previous results as required.

**0211-122:** The radio imaging of this object (Carilli et al. 1997) reveals a sharp bend in the radio jet ~1" to the E of the nucleus, which suggests that the E jet is interacting strongly with the ambient ISM. In addition to the quiescent component (FWHM = 470±60 km s$^{-1}$) already identified by VM2003, we detect in the vicinity of the continuum peak, a blueshifted, kinematically perturbed component ($v_s$ = -620±160 km s$^{-1}$, FWHM = 1290±160 km s$^{-1}$) to Lyα, CIV and HeII.

**0731+438:** The radio images of this object (Carilli et al. 1997) show no obvious morphological evidence to suggest that strong jet-gas interactions are taking place. VM2003 identified a region of enhanced, blueshifted line emission from the N EELR, and this is slightly broader than the emission in the rest of the EELR (~1000 compared to ~600 km s$^{-1}$).

**0828+193:** The NE radio jet is observed to undergo a series of bends (see the radio imaging of Carilli et al. 1997), one of which appears to occur within the HSBR (see figure 1 of VM2002). In Lyα, CIV and HeII, VM2002 detected a kinematically perturbed component (FWHM ~1200 km s$^{-1}$) confined by the radio structure, and detected a kinematically quiescent (FWHM <400 km s$^{-1}$) component. The latter component was detected across the entire extent of the line emission, including beyond the radio hotspots. We find that in the HSBR the NV and CIII] lines can also be decomposed into the perturbed and quiescent components. The relative contribution of the perturbed and quiescent component varies dramatically within the HSBR: to the SW of the nucleus, the quiescent component dominates the line emission, but to the NE, where the jet appears to bend, the perturbed component is dominant. These results suggest that jet-gas interactions are taking place in this object.

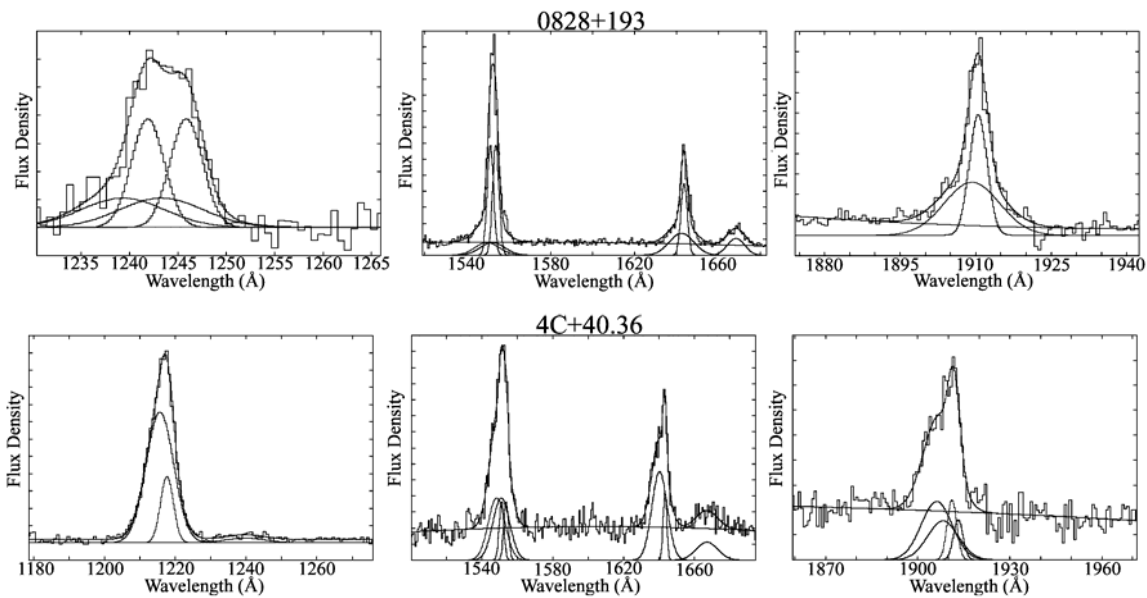

**Figure 2:** The fits to the perturbed and quiescent component in 4C+40.36 and 0828+193. The flux scale is arbitrary. The left-most column shows the fits to NV (in the case of 0828+193) or Lyα (for 4C+40.36); the central column shows the fits to CIV and HeII; the right-most column shows the fits to CIII]. It was not practicable to fit the perturbed and quiescent component for OIII] λ1663; however, the fit using a single kinematic component is shown for consistency. Similarly, for consistency the single Gaussian fit to NV is shown for 4C+40.36.

**Table 2.** Fit parameters for 4C+40.36 and 0828+193. The FWHM values are computed assuming that the object filled the slit. For the doublets, median wavelengths are quoted. The one-sigma uncertainties for the FWHM and wavelength measurements are about 100 km s$^{-1}$ and 4Å (~200 km s$^{-1}$), respectively. The velocity shift $v_s$ is given relative to the quiescent component of HeII in km s$^{-1}$, and has a typical one-sigma uncertainty of 200 km s$^{-1}$.

| 4C+40.36 | Quiescent | | | | Perturbed | | | |
|---|---|---|---|---|---|---|---|---|
| | Lyα | CIV | HeII | CIII] | Lyα | CIV | HeII | CIII] |
| FWHM (km s$^{-1}$) | 759 | 160 | <550 | 326 | 2131 | 2065 | 1816 | 1794 |
| Centre (Å) | 3975.6 | 5068.4 | 5365.8 | 6243.2 | 3969.1 | 5059.5 | 5355.8 | 6226.9 |
| $v_s$ (km s$^{-1}$) | -66 | 2 | 0 | 150 | -556 | -525 | -559 | -637 |
| 0828+193 | NV | CIV | HeII | CIII] | NV | CIV | HeII | CIII] |
| FWHM (km s$^{-1}$) | 652 | 581 | 376 | 462 | 2276 | 2276 | 1825 | 1977 |
| Centre (Å) | 4443.1 | 5545.2 | 5868.5 | 6824.4 | 4433.9 | 5535.0 | 5862.8 | 6819.9 |
| $v_s$ (km s$^{-1}$) | 281 | 108 | 0 | -16 | -341 | -444 | -291 | -213 |

**Table 3.** Line ratios for the perturbed and quiescent components for individual objects. The line ratios for 0943-242 were derived by isolating the perturbed and quiescent emission spatially (see §2). In the case of 0828+193 the ratios were derived by fitting the velocity profiles of each line of interest. For 4C+40.36, the CIV/HeII, CIV/CIII] and CIII]/HeII ratios were determined by fitting the velocity profiles, whereas the NV/CIV and NV/HeII ratios were derived by spatially isolating the perturbed and quiescent gas.

| Object | Quiescent Gas | | | | |
|---|---|---|---|---|---|
| | NV/CIV | NV/HeII | CIV/HeII | CIV/CIII] | CIII]/HeII |
| 4C+40.36 | 0.6±0.1 | 1.0±0.1 | 2.2±0.2 | 1.7±0.1 | 1.4±0.3 |
| 0828+193 | 0.37±0.03 | 1.0±0.1 | 2.0±0.2 | 2.5±0.2 | 0.85±0.08 |
| 0943-242 | 0.9±0.2 | 0.7±0.1 | 0.8±0.2 | 1.4±0.7 | 0.6±0.3 |
| Object | Perturbed Gas | | | | |
| | NV/CIV | NV/HeII | CIV/HeII | CIV/CIII] | CIII]/HeII |
| 4C+40.36 | 0.37±0.01 | 0.60±0.01 | 1.5±0.1 | 2.1±0.2 | 0.80±0.07 |
| 0828+193 | 0.32±0.03 | 0.49±0.04 | 1.5±0.1 | 3.6±0.03 | 0.45±0.04 |
| 0943-242 | 0.265±0.004 | 0.286±0.004 | 1.080±0.003 | 1.71±0.01 | 0.623±0.004 |

**0943-242:** In this object, VM2003 found that a perturbed component (FWHM ~1600 km s$^{-1}$) is confined within the spatial extent of the radio structure, and also that a kinematically quiescent component (FWHM ~500 km s$^{-1}$) extends across the entire object and far beyond the W radio hotspot. This suggests that jet-gas interactions are taking place in this object.

**4C-00.54:** The radio images (Carilli et al. 1997) and *HST* images (Pentericci et al. 1999, 2001) do not show any strong morphological evidence to suggest that jet-gas interactions are taking place. VM2003 detected only gas with quiescent kinematics (FWHM <472-800 km s$^{-1}$), again suggesting a lack of jet-gas interactions in this object.

**1558-003:** The kinematic decomposition carried out by VM2003 revealed the possible presence of a perturbed kinematic component, akin to that observed in other objects, though the fitting errors are somewhat large (see the discussion given by these authors). In any case, the HeII emission profile shows an excess of flux in its blue wing (see figure 1 of this paper). There is no clear evidence from the radio images (Pentericci et al. 2000b) or the long-slit spectroscopy (see e.g. figure 8 of VM2003) to suggest that jet-gas interactions are taking place.

**4C+40.36:** For this small (32 kpc) radio source, VM2003 found that the perturbed kinematic component (FWHM ~1600 km s$^{-1}$) is confined by the radio structure, while the quiescent component (FWHM ~600 km s$^{-1}$) extends across the entire observed nebulosity. The measurement of the He II surface luminosity of the quiescent kinematic component for this object (table 1) is rather uncertain due the apparent dominance of the perturbed component over the quiescent one. A perturbed component to [O III] λ5007 is detected by Egami et al. (2003), and has FWHM and velocity consistent with the fits of VM2003 and this paper. The Hα and [O III] λ5007 images of Egami et al. (2003) show that the high surface brightness line emission has a linear morphology, suggesting its spatial distribution may be the result of jet-gas interactions.

**4C+48.48:** Compelling evidence exists to suggest that the SW radio jet is interacting strongly with the ambient ISM. There is a bright hotspot and a ~90° clockwise bend in the radio jet, ~2" SW of the continuum centroid (see Chambers et al. 1996). It is also apparent that there are a number of morphological associations between the radio and UV-optical emission. In the combined K+H image of Carson et al. (2001) there is an arc, to which line emission contributes substantially, associated with the bright hotspot and which appears to follow the bend in the jet through ~90°. Moreover, the UV line and continuum emission is enhanced at this location (see e.g. figure 3 of VM2003). Complex kinematics are observed in the gas within the radio structures. VM2003 found the presence of a perturbed kinematic component (FWHM ~2200 km s$^{-1}$) within the radio structure, and a quiescent component (FWHM ~700 km s$^{-1}$) detected over the whole EELR. The gas associated with the bend in the radio jet has significantly lower ionization state than the gas beyond the radio structure and nearer the nucleus (Humphrey 2004; Humphrey et al. 2006, in preparation).

**2104-242:** The line emission in this object shows two bright clumps along the radio axis (e.g. McCarthy et al. 1990). Complex kinematics (i.e. multiple kinematic components) are observed in both the NE and SW regions, and both regions show an excess of flux in the blue wing of their velocity profiles (see e.g. Overzier et al. 2001 and VM2003). It is worth noting that this object was observed at higher spectral resolution (6Å) than for the rest of this sample (10-14Å), and therefore the blueshifted emission components seen in this object could mimic a single perturbed, blueshifted component when observed at lower resolution (see also the 2.8Å resolution Lyα spectrum presented by Pentericci et al. 2001).

There is a stark contrast in the properties of the NE and SW emission regions. The SW region shows a filamentary morphology in both Lyα and UV-optical continuum emission (Pentericci et al. 1999; 2001), and is aligned to within a few degrees of the radio axis. In this region the line emission shows highly complex kinematics, with several blueshifted components (Koekemoer et al. 1996; Pentericci et al. 2001; Overzier et al. 2001), and relatively low values for N V/He II and N V/C IV (Overzier et al. 2001). These properties bear a striking resemblance to the jet-gas interactions seen in 4C+48.48. The fact that there is no apparent enhancement of the radio emission associated with the SW region does not provide any strong argument against this notion, since many such cases have been reported in the literature (e.g. Koekemoer et al. 1996; Villar-Martin et al. 1998).

The emission to the NE of the radio core, on the other hand, is much more quiescent (Pentericci et al. 2001; Overzier et al. 2001) and shows a morphology reminiscent of an ionization cone (Pentericci et al. 2001). The values for N V/He II and N V/C IV ratios at this position are substantially larger than are observed for the filamentary structure (Overzier et al. 2001). Thus, the effects of jet-gas interaction on this region are far weaker than is the case for the SW region.

**Table 4.** [1] Name of the source  [2] Surface luminosity of the quiescent He II emission in the HSBR  [3] Surface luminosity of perturbed He II emission in the HSBR  [4] Luminosity of He II in the spatially integrated spectrum  [5] Linear polarization at 1500 Å  [6] Luminosity of the 1500 Å polarized continuum in the spatially integrated spectrum, which was obtained by multiplying the degree of polarization with the continuum luminosity at 1500 Å  [7] Luminosity of the young stellar population at 1500 Å, obtained by multiplying the fractional contribution from young stars by the continuum luminosity at 1500 Å.  NOTE – polarization measurements, continuum luminosities and the contribution from young stars were taken from Vernet et al. (2001).

| Object | $SL_q$ ($10^{40}$ erg s$^{-1}$ kpc$^{-2}$) | $SL_p$ ($10^{40}$ erg s$^{-1}$ kpc$^{-2}$) | $L_{HeII}$ ($10^{43}$ erg s$^{-1}$) | $P_{UV}$ (per cent) | $L_{Pol}$ ($10^{40}$ erg s$^{-1}$ Å$^{-1}$) | $L_{YSP}$ ($10^{40}$ erg s$^{-1}$ Å$^{-1}$) |
|---|---|---|---|---|---|---|
| [1] | [2] | [3] | [4] | [5] | [6] | [7] |
| 4C+40.36 | 0.5 (<5) | 9±3 | 1.4±0.4 | 7.3±1.2 | 1.2±0.4 | 6.2±3.5 |
| 0828+193 | 9±2 | 26±6 | 2.4±0.6 | 10.1±1.0 | 2.9±0.8 | 7.8±7.8 |
| 0943-242 | 11±4 | 10±3 | 1.5±0.4 | 6.6±0.9 | 1.7±0.8 | 13.7±8.1 |
| 1558-003 | 6±3 | 3±1 | 0.3±0.1 | - | - | - |
| 4C+48.48 | 9±2 | 5±1 | 1.0±0.3 | 8.4±1.5 | 1.5±0.5 | 7.5±4.6 |
| 0211-122 | 6±2 | 3±1 | 0.4±0.1 | 19.3±1.2 | 2.2±0.6 | 0.7±0.7 |
| 4C+23.56 | 9±3 | - | 0.3±0.1 | 15.3±2.0 | 2.2±0.7 | 2.6±2.6 |
| 4C-00.54 | 7±2 | - | 0.5±0.1 | 11.7±2.7 | 1.4±0.5 | 2.7±2.7 |
| 0731+438 | 13±4 | - | 0.6±0.2 | <2.4 | <0.4 | 10.5±2.3 |
| 2104-242 | 11±3 | - | 0.7±0.1 | - | - | - |

**4C+23.56:** The radio and optical images of this source (Chambers et al. 1996; Knopp & Chambers 1997) do not show any morphological evidence to suggest that jet-gas interactions are taking place. Indeed, the UV line and continuum emission shows a striking biconical morphology (Knopp & Chambers 1997) suggesting ionization and illumination by a hidden AGN. Only quiescent emission line gas is detected in this object (FWHM ~550 km s$^{-1}$; VM2003) and, therefore, this object shows no kinematic evidence for jet-gas interactions.

## 3.2 General results

The velocity profiles of the He II emission from the HSBR are shown, along with the single/double Gaussian fits, in figure 1. We show in Table 1 the kinematic properties and luminosity of the He II[‡] emission from the HSBR, along with the flux ratio between the perturbed and quiescent components (P/Q hereinafter). In all tables and figures, the peak of the quiescent component is taken to be the velocity zero. We also tabulate some salient properties of the radio structure. From Table 1 (column 7, showing the flux ratio of perturbed to quiescent gas P/Q) it is immediately apparent that some objects are dominated by the perturbed gas (4C+40.36, 0828+193), some have a mixture of perturbed and quiescent gas (0943-242, 4C+48.48 and 0211-122) and others are dominated by the quiescent gas (4C-00.54, 4C+23.56, 0731+438 and 2104-242). The perturbed component is always blueshifted with respect to the quiescent component; it is seen to be spatially extended in five objects (0828+193, 0943-242, 4C+40.36, 4C+48.48 and 2104-242); and it is detected on both sides of the nucleus in four objects (0828+193, 0943-242, 4C+40.36 and 2104-242); see figures 5-6 of VM2002 and figures 1-9 of VM2003.

It can also be seen from Table 1 that P/Q is strongly correlated with the global FWHM of He II, with a Spearman rank correlation coefficient of >94 per cent, or 100 per cent when the data are binned by a factor of 2 in P/Q or FWHM. In other words the global FWHM is determined by the balance between the perturbed and quiescent gases. Moreover, P/Q shows a striking relationship with the properties of the radio source: whereas radio sources which do not show any morphological evidence for hotspots or bends within the HSBR (4C-00.54, 4C+23.56, 0731+438 and 2104-242) have narrower He II and are dominated by the quiescent component, radio sources with radio hotspots or bends within the HSBR (4C+40.36, 0828+193, 0943-242, 4C+48.48 and 0211-122) have broader He II and a significant perturbed component, which often dominates the overall flux of the line.

Figure 2 shows for 4C+40.36 and 0828+193 the best-fits for the quiescent and perturbed components to C IV, He II, C III] and Lyα (4C+40.36) or N V (0828+193). Table 2 gives the parameters of these fits. Table 3 gives the line ratios derived for the perturbed and quiescent components of 4C+40.36, 0828+193 and 0943-242, the only sources for which both could be derived. Interestingly, in each of these three objects, the N V/C IV and N V/He II ratios are lower in the perturbed component than in the quiescent component. Furthermore, in both 4C+40.36 and 0828+193, the perturbed component has significantly lower C IV/He II and C III]/He II, and higher C IV/C III], than the quiescent component; it is unclear whether C IV/He II, C III]/He II or C IV/C III] differ significantly between the perturbed and quiescent gas of 0943-242, due to the relatively large errors. The N V/C IV ratio of the perturbed component is remarkably consistent from object to object, having a value of ~0.3; for this component, the other line ratios typically vary by a factor of ~2. The quiescent component shows a very large range in line ratios between different objects, and also within individual objects themselves. N V/C IV and N V/He II vary by a factor of ~10, C IV/He II and C III]/He II by a factor of ~3, and C IV/C III] by a factor of ~7 (see table 3 for 4C+40.36, 0828+193 and 0943-242; see Humphrey 2004 for the rest of the sample).

We present in table 4 the He II surface luminosity in the HSBR of the quiescent and perturbed components. We also show in this table the degree of polarization, the luminosity of the polarized continuum and the luminosity of the young stellar continuum, all of which are measured at 1500 Å and have been obtained from Vernet et al. (2001). It can be seen that the luminosity of young stars varies substantially throughout this sample (i.e. by about an order of magnitude) and, moreover, the luminosity of young stars is lower for objects with higher polarization, and is higher for objects with lower polarization; this is consistent with the possible anticorrelation of the luminosity of sub-mm emission, which Tadhunter et al. (2002) have argued is from starburst heated dust, with the degree of UV polarization (Reuland et al. 2004). Note that although the luminosity of the nebular continuum varies by about an order of magnitude throughout this sample, its contribution to the dilution of the

---
[‡] Note that using the kinematic properties and luminosity of other strong lines such as C IV or C III], in place of those of He II, does not affect significantly the results of this paper.

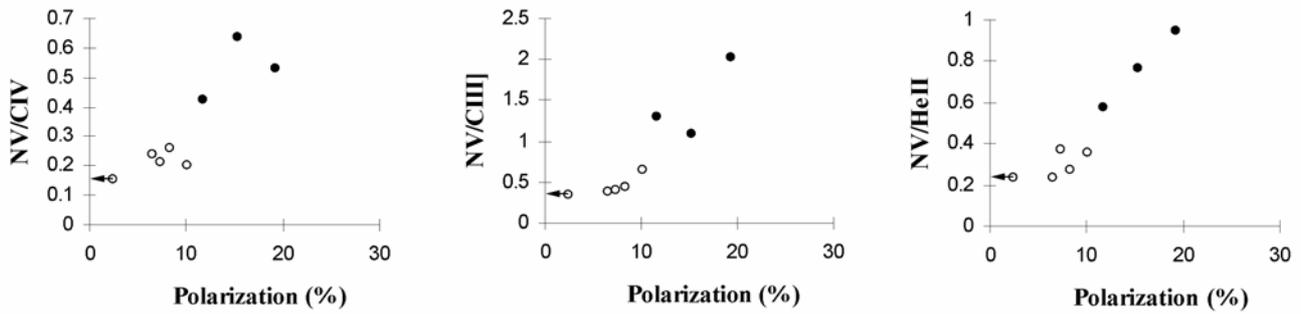

**Figure 3 (a):** The NV/CIV, NV/HeII, NV/CIII] and ratios against UV polarization. The three most polarized objects, 0211-122, 4C+23.56 and 4C-00.54, are represented by filled circles. Open circles represent 0731+438, 0828+193, 0943-242, 1558-003, 4C+40.36, 4C+48.48, and 2104-242; note that 1558-003 and 2104-242 do not have polarization measurements, and that the latter does not have a measurement for CIII]. In the interest of clarity, error bars are not shown; these are typically 10 per cent.

scattered light is insufficient to explain the variation in polarization within this sample (see Table 5 in Vernet et al. 2001). The quiescent HeII surface luminosity and the polarized continuum luminosity do not show any obvious relationship with each other; nor do they show any clear relationship with the surface luminosity of the perturbed HeII or the global HeII luminosity.

Vernet et al. (2001) have identified two interesting trends in the spatially integrated spectra of this dataset: the NV/CIV and NV/HeII ratios tend to be higher, and Lyα/CIV tends to be lower, in objects which have higher UV polarization. In order to further investigate these correlations, and their relationship to the ionization/excitation, radio and kinematic properties, line ratios involving NV, NIV] and Lyα are compared with other ratios formed from the strong lines NV, NIV], CIV, HeII and CIII], and with the UV continuum polarization, the radio size, and the luminosity and FWHM of HeII λ1640. These properties are taken from the spatially integrated spectroscopy presented by Vernet et al. (2001). We illustrate in figure 3 the apparent trends. In table 5 we compare the properties of the objects with higher UV polarization (namely 0211-122, 4C-00.54 and 4C+23.56) against those of the objects with lower UV polarization (0731+438, 0828+193, 0943-242, 4C+40.36 and 4C+48.48).

Objects which show lower UV polarization (open symbols in fig 3) tend to have lower NV/CIV, NV/HeII and NV/CIII] ratios [fig 3(a)], broader HeII and higher luminosity for HeII [fig 3(b)]. These objects are in general associated with smaller radio sources (<100 kpc) [fig 3(c)]. In addition, smaller radio sources tend to show lower values for NIV]/HeII and NIV]/CIII] [fig 3(a)]. See also table 5.

On the other hand, objects with higher UV polarization (closed symbols in fig 3) tend to show higher NV/CIV, NV/HeII and NV/CIII] ratios [fig 3(a)], and tend to have narrower and less luminous HeII emission [fig 3(b)]. Larger radio sources (>100 kpc) tend to show higher NV/CIV, NV/HeII, NV/CIII], NIV]/HeII and NIV]/CIII] ratios, and higher UV polarization [fig 3(c)]. Again, see table 5. 0731+438 is a notable exception to these trends: while this object shows narrow HeII and relatively low HeII luminosity, it has a relatively small radio source, relatively low polarisation and low NV and NIV] ratios.

**Table 5.** Mean properties of the objects with higher UV polarization (0211-122, 4C-00.54 and 4C+23.56) and objects with lower UV polarization (0731+438, 0828+193, 0943-242, 4C+40.36 and 4C+48.48). [1] Property [2] Mean value for high-polarization objects, 0211-122, 4C+23.56 and 4C-00.54 [3] Mean value for low-polarization objects. *4C-00.54 and 4C+23.56 only, because NIV] was not detected from 0211-122 (Vernet et al. 2001). NOTE – units are the same as for table 1 and table 4.

| Property [1] | High Pol. [2] | Low Pol. [3] |
|---|---|---|
| $P_{UV}$ | 15.4±1.2 | 7.0±0.5 |
| NV/CIV | 0.53±0.04 | 0.21±0.02 |
| NV/CIII] | 1.46±0.11 | 0.44±0.04 |
| NV/HeII | 0.76±0.05 | 0.29±0.03 |
| NIV]/HeII | 0.11±0.02* | 0.060±0.005 |
| NIV]/CIII] | 0.17±0.03* | 0.095±0.008 |
| $L_{HeII}$ | 0.40±0.06 | 1.4±0.2 |
| HeII FWHM | 700±100 | 1100±100 |
| Radio Size | 245 | 71 |
| $SL_q$ | 7±1 | 9±1 |
| $SL_p$ | 2±1 | 10±2 |
| $L_{Pol}$ | 1.9±0.3 | 1.5±0.3 |
| $L_{YSP}$ | 2±1 | 9±3 |

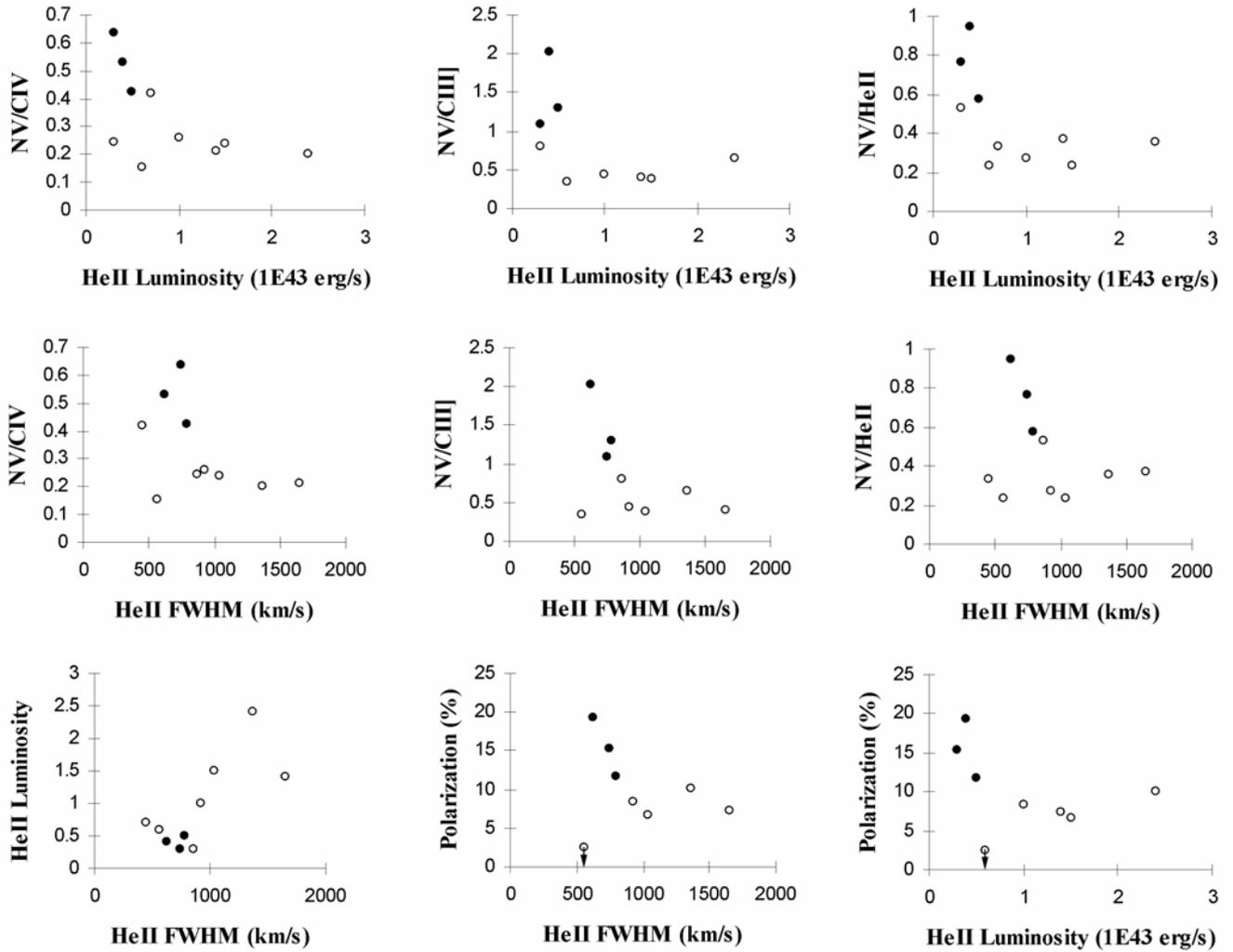

**Figure 3 (b):** NV/CIV, NV/HeII, NV/CIII] and UV polarisation against the luminosity and FWHM of HeII λ1640; also the FWHM against the luminosity of HeII λ1640. The symbols are as for figure 3 (a). Error bars are not shown; these are typically 10 per cent.

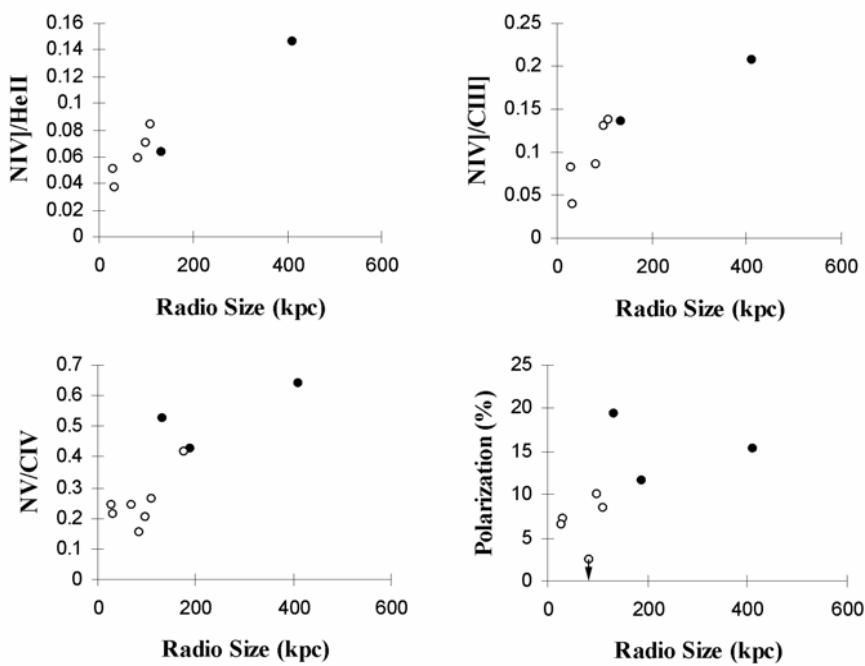

**Figure 3 (c):** NIV]/HeII, NIV]/CIII], NV/CIV and UV polarisation against radio size; the symbols are as for figure 3 (a), though it should be noted that the NIV] line has not been measured for 4C-00.54, 1558-003 and 2104-242. Also shown is the tendency for smaller radio sources to have lower NV/CIV and polarization. Error bars are not shown; these are typically 10 per cent.

Several of these apparent trends are reminiscent of the anticorrelation between ionization state and line width that has been reported for lower redshift radio galaxies that show clear signs of jet-gas interactions (Clark et al. 1997; Villar-Martín et al. 1999a; Best, Röttgering & Longair 2000); a weak correlation between polarization and radio size has also been reported by Solorzano-Iñarrea et al. (2004).

## 4. DISCUSSION

The nature of the kinematically quiescent and perturbed emission, their relation to each other, and finally, a scenario for the nature of jet-gas interactions, are in turn discussed.

Based on arguments involving energetics, VM2002 and VM2003 concluded that the quiescent line emission halo is photoionized by the hidden quasar, rather than being photo- or collisionally-ionized by shocks. The new results presented herein provide further support for this interpretation. A lower limit to ionizing photon flux through the volume sampled by our spatially integrated extraction apertures[§] from the AGN, can be extrapolated from the 1500Å polarized luminosities. We assume the SED of Mathews & Ferland (1987) and adopt a scattering efficiency factor of 1.5 per cent (Manzini & di Serego Alighieri 1996). A covering factor of unity is assumed for the ionized gas and for the scatterers; a lower covering factor would result in higher ionizing luminosities. For the purpose of this calculation, the EELR is assumed to be composed of ionization bounded, photoionized clouds. From the polarized continuum we obtain ionizing luminosities in the range Q ~ 0.4 - 4 × $10^{55}$ photons s$^{-1}$, broadly sufficient to power the observed quiescent line emission, which requires Q ~ 0.8 - 3.2 × $10^{55}$ photons s$^{-1}$. Moreover, the implied ionizing luminosities are sufficient to power the total line luminosity (i.e. quiescent plus perturbed components), which would require ~ 0.8 - 6.4 × $10^{55}$ photons s$^{-1}$. While this does not necessarily prove that the radiation field of the active nucleus is the only source of ionization for the EELR, it does show that the active nuclei of this sample are broadly capable of powering the line emission along the slit.

The fact that the perturbed gas is nearly always spatially extended and detected on both sides of the nucleus, and is always blueshifted with respect to the kinematically quiescent gas (figure 1, table 1), is rather a striking result. This is reminiscent of blue wings commonly seen in the nuclear narrow line regions of Seyfert Galaxies (Pelat, Alloin & Fosbury 1981; Wilson & Heckman 1984; Whittle 1985). The presence of blue asymmetries in HzRG velocity profiles has been reported for individual HzRG by a number of investigators (van Ojik et al. 1996; Pentericci et al. 2001; Maxfield et al. 2002; Jarvis et al. 2003). We propose that in our sample of high-redshift radio galaxies, jet-gas interactions induce a spherical or cylindrical expansion of material, with spatially extended dust obscuring the outflow on the far side of the radio axis. The considerable spatial extent of the blueshifted perturbed gas, typically ~35 kpc, implies that the dust in these objects is extended on at least a similar spatial scale. Entrainment processes appear to be required in order to explain the large line width (FWHM ~1550 km s$^{-1}$) of the perturbed gas observed in this sample (e.g. Villar-Martín et al. 1999a).

While outflows driven by the UV radiation field of the active nucleus (Binette 1998; van Bemmel et al. 2003) or winds and supernovae bubbles from a starburst (e.g. Heckman, Armus & Miley 1990) may provide an alternative means by which perturbed, blueshifted gas could be produced, in the light of the close association with hotspots or bends in the radio jets, such scenarios seem unlikely.

It is interesting to consider the fate of the kinematically perturbed gas that has been detected in many objects in this sample. Assuming the kinematically quiescent gas is in rotation, then its velocity shift across the EELR, typically ~450 km s$^{-1}$ (see table 2 of VM2003), implies a typical circular velocity of ~230 km s$^{-1}$ which in turn implies an escape velocity of $v_{esc} = \sqrt{2} v_{circ}$ ~320 km s$^{-1}$. The observed kinematic properties of the perturbed gas ($v_s$ of -200 to -650 km s$^{-1}$ relative to the quiescent gas and FWHM of 990 to 2200 km s$^{-1}$; see Table 1) then suggests that much of this gas is dynamically unbound and is in the process of being ejected from the host galaxy and its halo. Such outflows could inject enriched material into the haloes of the host galaxies and into the intergalactic or intracluster medium. Moreover, these outflows could be part of the solution to the so-called 'cooling-flow problem' (e.g. Binney & Tabor 1995; McNamara et al. 2000).

In explaining the ionization of the perturbed gas, we must explain why the perturbed gas varies significantly in surface luminosity from object to object, while the quiescent gas does not. The two most plausible scenarios are (i) additional gas is placed into the ionizing beam of the quasar by jet-gas interactions (Bremer, Fabian & Crawford 1997) and

---

[§] For each object the synthetic aperture used for the polarization measurement, typically ~8 × 40 kpc along the radio axis, was identical to that used for the extraction of the 1D total flux spectrum (see Vernet et al. 2001).

(ii) gas is ionized in situ by shocks generated by the jet-gas interaction. Also worthy of discussion is the significant difference in the line ratios between the perturbed and quiescent gases of 4C+40.36, 0828+193 and 0942-242: NV/HeII, NV/CIV and CIV/CIII] are lower, and CIV/HeII and CIII]/HeII are higher, in the perturbed gas than in the quiescent gas (table 3). This cannot readily be explained by photoionization with a difference in metallicity: lower metallicity gas is expected to have lower NV/HeII, NV/CIV, CIV/HeII and CIII]/HeII, and higher CIV/CIII] (see Vernet et al. 2001), and vice-versa. A more plausible alternative is that the perturbed gas has a lower ionization state than the quiescent gas. In a future paper (Humphrey et al. 2006, in preparation) we investigate further the ionization and physical conditions of the EELR.

We now propose a scenario to explain the apparent interrelationship between ionization state, the luminosity of HeII, the FWHM of HeII, the UV polarization and the projected extent of the radio emission. This builds on scenarios suggested by Clark et al. (1997, 1998), Villar-Martín et al. (1999a) and Best, Röttgering & Longair (2000). Intrinsic to each HzRG in this sample is a luminous AGN, and this is responsible for the polarized UV continuum and also for the ionization of the kinematically quiescent gas. The tendency for objects with broader HeII to show lower NV/CIV, NV/HeII and NV/CIII] and more luminous HeII [fig 3(b)], is due to a variation between objects in the admixture of the perturbed, relatively low-ionization and quiescent, relatively high-ionization gases: as the contribution from perturbed ionized gas increases, with the luminosity of the quiescent gas remaining roughly constant, the global FWHM and luminosity of HeII also increase, while the mean ionization state decreases[**].

The tendency for smaller radio sources to have lower ionization state, as measured by the NV and NIV] ratios [fig 3(c)], than larger radio sources is produced by a combination of two effects. Firstly, as proposed by Best, Röttgering & Longair (2000), smaller radio sources are more likely to be interacting strongly with ISM of the host galaxy. This means that the relatively low-ionization, kinematically perturbed gas resulting from jet-gas interactions is more likely to be present in smaller radio sources. The second effect is that the radio jet/structure propagates more slowly, or is more effectively confined, in denser ISM (e.g. Swarup & Banhatti 1981); the corollary of this is that in smaller radio sources, the AGN photoionized gas is likely to have lower ionization parameter.

The tendency for the degree of UV polarization to be lower in objects with smaller radio sources and lower NV/CIV, NV/HeII and NV/CIII] ratios, and higher in objects with larger radio sources and higher NV/CIV, NV/HeII and NV/CIII] ratios (table 5), is another interesting result that must be explained within the framework of our scenario: objects which have smaller radio sources, and hence lower ionization state due to their stronger jet-gas interactions, tend to contain a more luminous young stellar population. Possible reasons for this include (i) jet-induced star formation (e.g. McCarthy et al. 1987; Chambers, Miley & van Breugel 1987), (ii) the observed radio-loud phase is triggered during a starburst/merger event (e.g. Smith & Heckman 1989) and (iii) feedback from the radio jet and active nucleus halts or hinders further star formation activity (e.g. Sazonov et al. 2005), and the pre-existing young stellar population then fades significantly on a timescale similar to that of the radio source ($10^7$ yr; e.g. Best, Longair & Röttgering 1996). Based on the data presented herein, it would be difficult to distinguish between these three possibilities.

## 5. SUMMARY

An investigation has been carried out into the nature of the interaction between the radio source and the ambient ISM, considering kinematic, energetic and ionization information for a sample of 11 radio galaxies at 2.3<z<2.9.

We have found that the kinematically perturbed ionized gas, which was shown in previous studies to be within the radio structures in six objects from this sample (VM2002; VM2003), is always blueshifted with respect to the kinematically quiescent ionized gas. This perturbed gas is spatially extended in five out of six objects, and is detected on both sides of the nucleus in four out of six objects. We propose that the perturbed gas is part of a jet-induced outflow, with dust obscuring the outflowing gas that lies on the far side of the object, receding from the observer. The spatial extent of the blueshifted perturbed gas, typically ~35 kpc, implies that the dust in these objects is spatially extended on similar spatial scales. We also suggest that in each of 4C+40.36, 0828+193 and 0943-242, the perturbed gas has a lower ionization state than the quiescent gas.

We have extrapolated, from the polarized continuum luminosity, the ionizing photon flux Q through the volume sampled by our extraction apertures (~8 × 40 kpc along the radio axis) and have compared this against the ionizing luminosity required to power the UV recombination line emission in these same apertures: both values for Q are broadly consistent.

---

[**] The variation in NV/CIV, NV/HeII, NV/CIII], NIV]/HeII and NIV]/CIII] is predominantly the result of a variation in ionization state throughout this sample (Humphrey 2004; Humphrey et al. 2006 in preparation).

Objects with lower UV polarization generally have lower NV/CIV, NV/HeII and NV/CIII] ratios, broader HeII and higher luminosity for HeII. Similarly, objects which show higher UV polarization tend to show higher NV/CIV, NV/HeII and NV/CIII], and generally have narrower and less luminous HeII emission. Larger radio sources (>100 kpc) generally show higher NV/CIV, NV/HeII, NV/CIII], NIV]/HeII and NIV]/CIII] ratios, and also higher UV polarization. This we propose is the result of a variation in the admixture of relatively high ionization, quiescent gas and relatively low ionization, perturbed gas, with the absolute luminosity of the quiescent gas remaining relatively constant throughout the sample.

Smaller radio sources generally show lower NV/CIV, NV/HeII, NV/CIII], NIV]/HeII and NIV]/CIII] ratios, while larger radio sources tend to show higher values for the NV/CIV, NV/HeII, NV/CIII], NIV]/HeII and NIV]/CIII] ratios. This is likely to be due to a combination of two effects: firstly, smaller radio sources are more likely to be undergoing strong jet-gas interactions, meaning that these objects are more likely to contain relatively low-ionization, kinematically perturbed gas resulting from jet-gas interactions. Secondly, the radio jet/structure propagates more slowly, or is more effectively confined, in denser ISM. The consequence of this is that in smaller radio sources, the denser ISM is likely to result in the AGN photoionized gas having lower ionization parameter/state. In addition, smaller radio sources tend to show lower UV polarization than do larger radio sources. We propose that the smaller radio sources tend to contain more luminous young stellar populations than the larger radio sources.

We have shown that although interactions between the radio source and the ambient ISM can have a profound impact on the properties of powerful high redshift radio galaxies, the influence of these interactions can be isolated cleanly. This will allow the properties and physical conditions of the emission line gas to be determined in a way that is free from the degeneracy between the current shock and photoionization models.

## ACKNOWLEDGEMENTS

Based on observations collected at the W. M. Keck Observatory, at the European Southern Observatory, and at the NRAO Very Large Array. AH acknowledges generous support from a PPARC studentship, a University of Hertfordshire post-doctoral research assistantship and a UNAM postdoctoral research fellowship. The work of MVM has been supported by the Spanish Ministerio de Educatión y Ciencia and the Junta de Andalucía through the grants AYA2004-02703 and TIC-114 respectively. We would like to thank Chris Carilli for providing the VLA radio maps, and thank Laura Pentericci and the Leiden HzRG group for the VLT spectrum for 2104-242. AH thanks Clive Tadhunter, Andy Robinson, Martin Hardcastle and Nial Tanvir for stimulating discussions which helped to improve this paper. We would also like to thank the anonymous referee for useful comments.